\documentclass[fleqn,usenatbib]{mnras}

\usepackage{newtxtext,newtxmath}

\usepackage[T1]{fontenc}

\DeclareRobustCommand{\VAN}[3]{#2}
\let\VANthebibliography\thebibliography
\def\thebibliography{\DeclareRobustCommand{\VAN}[3]{##3}\VANthebibliography}

\usepackage{graphicx}	
\usepackage{amsmath}	

\title[Waiting in the wings]{Waiting in the wings: extended emission GRBs from high latitude emission and jet expansion}

\author[I. Worssam et al.]{
I. Worssam,$^{1,2}$\thanks{E-mail: ixw352@student.bham.ac.uk}
B. P. Gompertz,$^{1,2}$
H. J. van Eerten$^{3}$
\\
$^{1}$School of Physics and Astronomy, University of Birmingham, Edgbaston, Birmingham, B15 2TT, UK\\
$^{2}$Institute for Gravitational Wave Astronomy, University of Birmingham, Edgbaston, Birmingham, B15 2TT, UK\\
$^{3}$Department of Physics, University of Bath, Claverton Down, Bath BA2 7AY, UK
}

\date{Accepted XXX. Received YYY; in original form ZZZ}

\pubyear{\the\year{}}

\begin{document}
\label{firstpage}
\pagerange{\pageref{firstpage}--\pageref{lastpage}}
\maketitle

\begin{abstract}

The process powering extended emission (EE) Gamma Ray-Bursts (GRBs) from compact object mergers is a debated problem due to their observed prompt emission timescales of 10 to 100\,s, far in excess of the expected accretion timescales following a binary neutron star (BNS) merger. Here, we present an analytical thin shell structured jet model designed to alleviate this tension. We show that a changing jet structure in which early emission is produced by a high Lorentz factor narrow, structured jet and later emission comes from a wider, lower Lorentz factor tophat-like jet can successfully reproduce the EE GRB phenomenology. All emitting shells are powered by a $0.1$\,$M_{\odot}$ accretion torus and launched within a time window of 2\,s after merger. Spatially-resolved opacity checks confirm that our emission sites remain optically thin, in part due to the spectral softness associated with EE. We present light curves from our model compared to representative EE bursts detected by \emph{Swift}: GRBs 211211A, 211227A and 060614. While our light curves provide good matches to the data, our spectra do not evolve fast enough when compared to the well-measured spectral evolution observed in GRB 211211A. We discuss an update to the model incorporating inherent spectral evolution in the shells as future work. We also show a parameter exploration of the model and identify a parameter range within which standard short GRB-like light curves could be produced, showcasing the ability of the model to produce short bursts with and without EE.


\end{abstract}




\section{Introduction}

Gamma-ray bursts (GRBs) are one of the brightest events in the Universe \citep{brightGRBs, fireballmodel}. They consist of relativistic jets driven by two different progenitors: compact object mergers and collapsing massive stars (collapsars) \citep{Eichler1989, collapsarGRBs}. The GRB signal has two parts: the prompt emission, which originates from internal processes within the jet \citep{internalcollisionsmodel, MVT_internalshocks, Daigne_internalshocks}, and the afterglow, which is the consequence of the interaction between the jet and the surrounding medium \citep{afterglowmodel, ReesMeszaros1997_afterglow, Waxman_afterglow_externalinteraction}. The prompt emission duration varies from millisecond pulses to activity lasting for hundreds of seconds and consists of stochastic pulses of emission predominantly observed in the gamma- and X-ray bands \citep{Sari1996}. The GRB duration is classified by the parameter $T_{90}$: a measurement of the time over which the middle 90\% of the prompt fluence is received \citep[cutting the first and last 5\% in order to remove potential biases from fading tails of emission;][]{2s_split}. Afterglows rise later, after the jet has swept up enough mass from the external medium to be decelerated and for shocks to form. They are typically observed for days to weeks across X-ray, optical and radio frequencies \citep{Waxman_afterglowtimescale, Sari1998}. 

Historically, GRBs with prompt emission lasting less than two seconds are deemed short GRBs, whereas those lasting longer than two seconds are termed long GRBs \citep{2s_split}. This is an empirically determined classification formed from a bimodal distribution in the T$_{90}$ of GRBs seen in observational data. The separation point between the two distributions is detector dependent \citep[e.g.][]{detectordependent_split}, lending some inherent uncertainty to this classification method. Long GRBs are also observed to have positive spectral lags \citep[where low energy photons are received delayed with respect to high energy photons;][]{long_spectrallags, long_spectrallags2} whereas short GRBs tend to exhibit negligible spectral lags \citep{EE_GRBs, short_spectrallags}. Their host galaxies are also observed to differ, with short GRBs preferring quiescent galaxies and residing at larger offsets from the galactic centre \citep{host_offsets, offset+inactivehosts} compared to long GRBs that are more commonly found in active, star forming galaxies \citep{long_hosts, long_hosts2}. Long GRBs have been found to correlate with type Ic broad line supernovae \citep{SN-GRBs2, SN-GRBs3, SN-GRBs} with the first connection of an exceptionally bright supernova 1998bw with GRB 980425 \citep{SN1998bw}. This correlation of supernovae and long GRBs leads to the association of these GRBs to collapsing stars.

The shorter timescales of short GRBs align with the compact object merger scenario. Indeed, the detection of gravitational waves associated with short GRB 170817A \citep{GW170817} confirmed the source as a binary neutron star (BNS) merger. A BNS merger can also produce a kilonova: the radioactive glow of elements produced through $r$-process nucleosynthesis. Kilonovae can also be used therefore to identify merger driven GRBs. However, they are faint transients with a fast decay time and hence hard to catch. Consequently, the population of known kilonovae is very small \citep{130603B_Tanvir, 130603BKN, 060614KN, 080503KN, 050709KN, 170817_EM_LIGOpaper, 170817_2, 170817_3, 170817_4, 170817KN, 160821BKN, 150101B, 070809KN, 211211A_kilonova_Troja, 211211A_kilonova_Jillian, 230307A_KN2, 230307AKN, 230307A_KN3, 230307A_Gillanders}.


However, the ostensibly long GRB 211211A has exposed a tension in this classification scheme. Despite its long $T_{90}$ of 51\,s \citep{211211AGCN}, no accompanying supernova was detected. Instead, a kilonova was detected \citep{211211A_kilonova_Jillian, 211211A_kilonova_Troja}, marking the GRB as merger powered. This confirmed long-held suspicions of the existence of long duration, merger powered GRBs motivated by other long GRBs lacking a supernova detection (for example, GRBs 050724 \citep{050724}, 051109B \citep{051109B}, 111005A \citep{111005A}, 060505 \citep{060505}, 211227A \citep{211227A} and 191019A \citep{191019A}) allowing them to be considered as potential merger candidates. 

There is now a population of $\sim$ ten GRBs with associated candidate kilonovae. Two of them are spectroscopically confirmed: GRB 170817A \citep{170817KN} and GRB 230307A \citep{230307AKN}. An additional two have very well-sampled infrared excesses that are highly consistent with the GW170817 kilonova: GRB 160821B \citep{160821BKN} and GRB 211211A \citep{211211AKN}. Three of these candidate kilonovae are associated with long GRBs: GRB 060614 \citep{060614KN}, GRB 211211A \citep{211211AKN} and GRB 230307A \citep{230307AKN}. These observations strongly suggest a merger origin is capable of producing a long GRB as well.

In a BNS merger tidal forces strip the outer layers off the neutron stars and an accretion torus is formed around the resulting black hole \citep{accretion_timescale}. The in-fall of this material onto the black hole provides the engine that powers the GRB. In general, the accretion time is of the order of seconds \citep{accretion_timescale2, accretion_timescale}, consistent with the short GRB population. However, this presents a problem when trying to explain the longer duration of these long GRBs of merger origin (long mergers).

The prompt emission of these long mergers appears to follow a common shape. The light curves of these GRBs appear to follow a pattern of an initial pulse complex (IPC), which consists of stochastic, spiky emission, followed by a more plateaued section, at lower energies, termed extended emission (EE) \citep{EE_GRBs, Perley2009,BensMagnetar,Dichiara2021, machinelearning_EErefs, EE_MVT}. 

The length of the IPC in this long merger population varies. Some such as GRB 211211A have long IPCs of tens of seconds \citep{211211AGCN} whereas others \citep[such as GRB 211227A;][]{211227AGCN} have short IPCs but a long overall $T_{90}$ with the addition of the EE plateau. There is a potential that all merger GRBs have this EE signal but for the `standard' short GRBs, where extended emission is not seen, it instead exists at lower energies (outside the bandpass of gamma-ray detectors that observe the prompt emission). This would indicate that potentially the EE shape is diagnostic of the merger progenitor \citep{Perley2009, EEcontinuum}. The recent addition of Einstein Probe (EP) and SVOM \citep{SVOMmission, EPmission} to the sky bring hope that extended emission from short GRBs might be captured with these lower energy telescopes that observe primarily in the X-ray band, allowing this to be investigated.

There are already many competing theories in the literature attempting to explain the presence of long merger GRBs. Broadly these theories can be split into two categories: those that invoke an additional energy injection parameter and those that attempt to extend the accretion timescale. 

A few examples invoking additional energy parameters include the formation of a millisecond magnetar before collapse to a black hole. In this model the rotational spin of the magnetar (a heavily magnetised, fast-spinning neutron star) releases additional energy into the jet hence allowing a longer GRB \citep{Metzger_Magnetar, Bucciantini_magnetar, Ben_expaccretion}. 

While BNS mergers are established GRB progenitors, GRBs may also arise from black hole - neutron star (BH-NS) mergers \citep{NS-BH}, although this is unconfirmed observationally. Outwards thermal pressure from the kilonova can clear a path for later fallback accretion to power the EE from marginally bound material that is still gravitationally bound due to the larger mass of the remnant \citep{NS-BH, Troja_NSBH, NSBH_KNfallback, Ben_NSBH}. 

Without invoking additional energy parameters, \cite{magnetic_braking} proposes a large magnetic flux build-up in the accretion disk, impeding the in-fall of accretion onto the black hole, slowing the accretion process and hence extending the timescale of the GRB emission. While this does extend the timescale of accretion it spreads the accreting material out over a longer passage of time and hence the energy budget per second decreases, lowering the instantaneous flux output of the GRB. 

\cite{MAD} suggest a massive accretion disk can naturally give rise to long merger GRBs. They propose the development of a global toroidal magnetic field that contributes to an increase in the jet launch efficiency, causing a brightening effect that extends the timescale of emission even as the accretion rate declines.


\cite{vv_2jet} argue a 2-component jet model consisting of a neutrino-annihilation-powered wider jet and a narrower, Blandford-Znajek \citep{BZpaper} powered jet can produce GRBs with and without EE dependent on the viewing angle of the observer. 

The expected short-duration of neutron star merger-produced GRBs is an order-of-magnitude estimate based on the characteristic lifetime of the central engine. All observable prompt emission is not necessarily confined to this timescale. Angular structure within the relativistic outflow can cause parts of the jet to emit later, particularly when allowing for a difference in maximum photon energy and total energy output between early and later stages of prompt emission. Consequently, a relativistic jet launched over a short period may still produce observable emission on substantially longer timescales. It therefore remains of interest to explore the parameter space of a generic model for prompt emission from relativistic jets in detail to determine whether late prompt emission can arise under certain conditions.

Our aim in this work is to reconcile the observed long duration of long merger GRBs with the short expected accretion timescale following BNS mergers, while reproducing some of the features of the light curves such as the EE plateau. We use a simple accretion torus model with a 0.1 $M_\odot$ energy reservoir. Utilising a launch time distribution, a lateral jet structure and modelling high latitude emission (HLE), our model extends the observed timescale and is able to reproduce IPC plus EE-like light curves.

This paper is organised as follows. In Section \ref{sec:model setup} we describe the structure of the model, including jet launch, high latitude emission, jet expansion and emission production. In Section \ref{sec:lightcurves} we  present our canonical model light curve, compare it to data from representative EE bursts and explore the free parameter dependencies. In Section \ref{sec:discussion} we discuss the feasibility and implications of the assumptions within our model. Our conclusions are presented in Section~\ref{sec:conclusions}.

\section{Model Setup}
\label{sec:model setup}

\subsection{Jet Launch}
\label{subsec: Jet launch}

We model the jet outflow as a series of discrete, thin (infinitesimal width) shells of ejecta launched by a central engine, powered by the energy reservoir of an accretion torus formed during the merger. In a BNS merger, accretion is expected to last for only seconds \citep{torus_sims, MAD}, often taken as less than two seconds, corresponding to the duration of short GRBs \citep{accretion_timescale}. Therefore, in order to ensure that our model is consistent with constraints from the accretion timescale, we launch all our shells over a two second time period. Once all the shells have launched, at the two second mark, the engine is no longer active. 




The launched mass initially rises, similar to the increasing jet launch efficiency employed in the MAD model \cite{MAD}, before following an exponential decline, chosen to mimic the expected accretion profile of the central engine \citep{Ritter_expaccretion, expaccretion_KNwinds, Ben_expaccretion}.

The mass per shell is expressed as 

\begin{equation}
    M_i = \eta_{\rm launch} M_{\rm torus} \frac{\exp(-t_{{\rm launch}, i} / \kappa)}{\sum \exp(-t_{{\rm launch}, i} / \kappa)},
    \label{eqn:exp distribution}
\end{equation}

\noindent where $M_i$ is the mass of each shell, $\eta_{\rm launch}$ is an efficiency parameter rising from 5x$10^{-4}$ to 0.05 over a time period of 0.4\,s, $M_{\rm torus}$ is the total mass of the accretion torus, $t_{{\rm launch}, i}$ is the time each shell is launched ($t_{{\rm launch}, i} < 2 s$) and $\kappa$ is the turnover time of the exponential function where $\kappa < 2\,s$. 

Each shell is divided into emission regions along its lateral width ($d\theta$) and $M_i$ is distributed evenly across these emission regions per shell.

Alongside this mass profile, we assume an exponential distribution of launch Lorentz factors, matching that of the shell masses above. Later shells therefore have lower Lorentz factors when there is less material accreting on to the black hole. 

We also include a lateral distribution of Lorentz factors across each shell, creating a structured jet \citep{170817_structuredjets, gavin_structuredjets, OConnor_structuredjets}. We parameterize the lateral Lorentz factor distribution as an energetic core embedded in slower material which forms the wings of the jet. We model a smooth and gradual decrease in Lorentz factor with angle as opposed to a discrete jump between a core and cocoon. We choose a Gaussian distribution to match theoretical assumptions \citep[e.g.][]{gavin_structuredjets,gaussianjet_model} and observations \citep[e.g.][]{170817_gaussianjet_data,170817_gaussianjetmodel_+baryonloading}, with the maximum Lorentz factor at the head of the shell and a symmetric decrease along the wings.  

Our overall Lorentz factor distribution is then:

\begin{equation}
    \Gamma_{i, \theta} = \Gamma_0 \exp \left[ -t_{{\rm launch}, i} / \kappa \right] \times \exp \left[ - \theta / 2 \theta_c^2 \right],
    \label{eqn:lorentz factor distribution}
\end{equation}

\noindent where $\Gamma_{i, \theta}$ is the Lorentz factor of each emission region (each angular element $\theta$ on shell $i$), $\Gamma_0$ is the maximum Lorentz factor, $\theta$ is the angle from the central axis and $\theta_c$ is the angular size of the energetic core. 

Combining our Lorentz factor and mass distributions, our total accretion to jet energy efficiency is 0.7\% (with a 30\% emission efficiency) which is just below the range derived by \cite{neutrinojets_efficiency} and less than the maximum determined for long GRBs by \cite{1.5efficiency}.


We employ an expansion of the jet by launching wider shells after a designated time $t_{\rm trigger}$. We determine $t_{\rm trigger}$ from the time at which the first shell below a pre-defined Lorentz factor threshold (where we use the Lorentz factor at the head of the shell) is launched by inverting the exponential function

\begin{equation}
    t_{\rm trigger} = -\kappa \ln(\Gamma_t / \Gamma_0)
    \label{eqn:trigger time}
\end{equation}

\noindent where $\Gamma_t$ is the trigger Lorentz factor.

Shells launched at $t_{\rm launch} < t_{\rm trigger}$ are narrower with an angular size of $\theta_{\rm narrow}$ while shells launched at $t_{\rm launch} > t_{\rm trigger}$ are wider with a width of $\theta_{\rm wider}$. To create a gradual widening of the jet, $\theta_{\rm wider}$ gradually increases from $\theta_{\rm narrow}$ to a maximum value. This is set through a smoothing parameter where a higher smoothness corresponds to a larger number of shells over which the jet expands. We treat the expansion as a stretching of the jet and alongside it also increase the relative size of the core. This reduces the steepness of the lateral Lorentz factor distribution at late times. 

\begin{figure}
    \centering
    \includegraphics[width=\columnwidth]{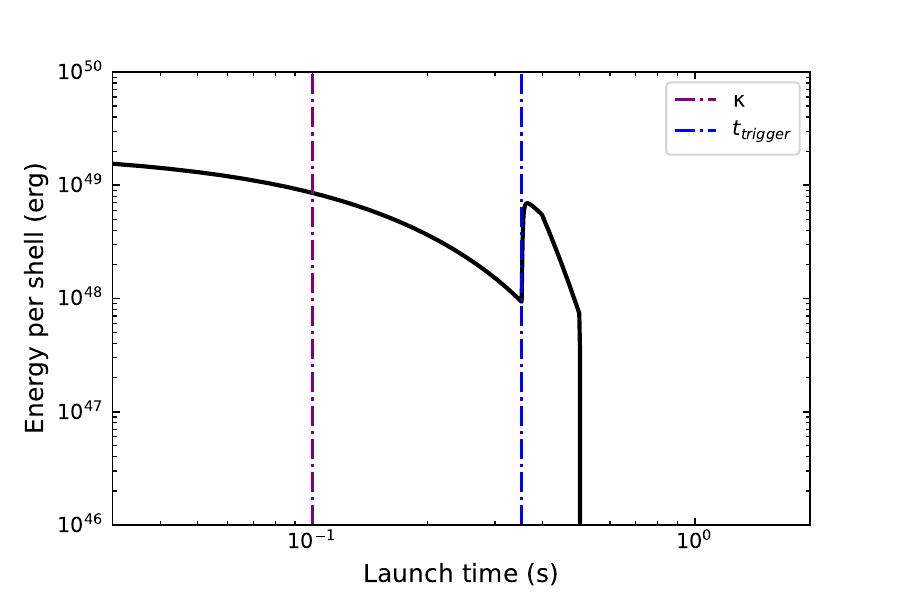}
    \caption{The launched energy profile of the shells, energy per shell over the launch time (lifetime of the central engine). The blue dotted line shows the time the jet expansion begins and the purple dotted line marks the turnover time of the exponential launch profiles.}
    \label{fig:energy profile}
\end{figure}

Figure~\ref{fig:energy profile} shows the energy per shell over the 2\,s launch time. The deviation from the exponential decay at $t_{\rm trigger}$ corresponds to the onset of the release of the wider shells and is a consequence of the increase in emitting surface. This deviation from a smooth exponential function is still within the 2\,s time window for accretion. Therefore, we are not invoking any additional energy sources beyond the energy reservoir provided by the accretion torus. 

We truncate the emission below a Lorentz factor of two, using this as the limit of relativistic ejecta \citep{internalcollisionsmodel, gamma2_limit}. The sharp drop in Figure~\ref{fig:energy profile} is merely an indication of where the Lorentz factor of the shells falls below 2 and hence we no longer track the emission. 



\subsection{Reference Frames and Arrival Times}
\label{subsec:Dynamics}

Due to the velocity structure across our shells (each emission region has a unique Lorentz factor defined by Equation \ref{eqn:lorentz factor distribution}), there is a co-moving frame associated with each emission region per shell. We denote these frames with a dash (e.g. $t'$).  

There is also an engine frame (denoted without a dash) in which the central engine is at rest and an observer frame ($t_{\rm obs}$) that accounts for arrival time to a distant observer (where we neglect cosmological redshift, working in luminosity rather than flux space). 

We calculate the time of emission of each shell based on a pre-defined radius at which the head of each shell emits. We investigate two scenarios: a singular emission radius where the head of every shell emits at the same radius, and a linear spread across a small range where the heads of the later shells emit at larger radii \citep[an expanding emission radius, cf.][]{emissionradius_expanding2, emissionradius_expanding}. 

The time of emission is calculated using 

\begin{equation}
    t_{e, i} = \frac{R_i}{\beta_{i} c} + t_{{\rm launch}, i},
    \label{eqn:t_e,head}
\end{equation}
where $t_{e, i}$ is the time at which photons at the head of shell $i$ are emitted (as measured in the engine frame), $t_{{\rm launch}, i}$ is the time the shell was launched from the central engine, $\beta_{i} = \frac{v_{i}}{c}$ where $v_{i}$ is the velocity at the head of the shell and $R_i$ is the emission radius. 

We transform this emission time into the co-moving frame of the emitting region at the head of each shell using 

\begin{equation}
    t'_{e, i} = \frac{1}{\Gamma_{i}} t_{e, i},
    \label{eqn:fluid to eng}
\end{equation}
where $t'_{e, i}$ is the emission time of the head of the shell in the co-moving frame of that emission region and $\Gamma_{i}$ is the Lorentz factor of the emitting region. 

We then assume that in the co-moving frame of the head of each shell, every emission region on the shell emits simultaneously. This allows us to transform back into the engine frame and compute a time of emission for each lateral emission region per shell

\begin{equation}
    t_{e, i, \theta} = \Gamma_{i, \theta} t'_{e, i}
\end{equation}
where $\Gamma_{i, \theta}$ is the Lorentz factor of each lateral emission region. 

From here we can calculate an emission radius for each lateral emission region using 

\begin{equation}
    R_{i, \theta} = \beta_{i, \theta} c t_{e, i, \theta},
    \label{eqn:radius}
\end{equation}
where $\beta_{i, \theta} = \frac{v_{i, \theta}}{c}$ and $v_{i, \theta}$ is the velocity of each emitting region (using the same definition of radius as in Equation~\ref{eqn:t_e,head}).

\begin{figure}
    \centering
    \includegraphics[width=\columnwidth]{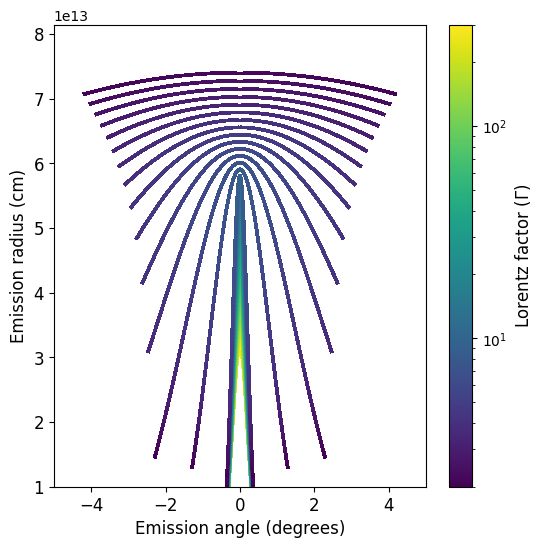}
    \caption{Diagram of a small subset of individual shells within the jet. Each line is a launched shell with emission regions plotted as angle from central axis against their emission radius. The colour bar displays the Lorentz factor across the shells. A minimum cutoff of $\Gamma = 2$ is applied, imposing this as the limit of relativistic material \citep{internalcollisionsmodel, gamma2_limit}. The expansion of the jet can be seen as the shells increase in width (x-axis) and become more uniform in Lorentz factor (colour gradient) compared to the Gaussian, earlier and narrow shells.}
    \label{fig:JetShape}
\end{figure}

Figure~\ref{fig:JetShape} shows a small subset of individual shells within the jet (far fewer than used in the presented light curves). Each shell is plotted as angle from the central axis ($\theta$) against emission radii (calculated using Equation~\ref{eqn:radius}). The lateral structure of the jet can be seen from the colour gradient that depicts the Lorentz factor. An initially narrow and higher Lorentz factor core expands into slower and wider, more uniform shells as the jet transitions from a Gaussian to a tophat structure.  

Photons emitted simultaneously from each lateral emission region along a shell reach the observer with a spread of arrival times quantified by \citep[cf.][]{t_obs_eqn, Sari1998}

\begin{equation}
    t_{{\rm obs}, i, \theta} = t_{e, i, \theta}(1 - \beta_{i, \theta} \cos\theta)
    \label{eqn:t_obs}
\end{equation}

where $t_{{\rm obs}, i, \theta}$ is the time emission from each lateral emission region per shell is received (in the observer frame). This expression combines the geometry of the shells, where emission from the wings has further to travel to reach the observer, with Doppler boosting 

\begin{equation}
    D = 1/\gamma(1 - \beta\cos\theta).
    \label{eqn:Doppler}
\end{equation}

\noindent which is angular dependent and hence emission from wider angles is less boosted. 

At each emission region the angle to the observer is compared to the relativistic beaming angle, $\phi_{i, \theta} = 1/\Gamma_{i,\theta}$. Only emission where the latter is larger ($\phi > \theta$) contributes to the light curve. In reality, the beaming angle is not a hard cutoff as approximated here for ease of computation. We therefore tested a lower cutoff of $\phi_{i, \theta} = 3/\Gamma_{i,\theta}$ to determine the impact of this assumption. No noticeable differences to our light curves were found. 





The individual pulses from each shell overlap, with emission from the head of the subsequent shell reaching the observer before light from the edge of the wings of the preceding shell. Hence, the final light curve is a composite sum of the individual shell pulses. We therefore calculate equal arrival time surfaces (EATS) to create a composite light curve from our many shells. 

The angular resolution of the model is set by the number of emission regions ($d\theta$) across the lateral width of each shell. We use a resolution of 10000 emission regions, the minimum for which the final light curve is not impacted. 

To calculate the EATSs we define a series of time bins. The width of these time bins is the time resolution of our final light curve. We ensure the time resolution is not less than the time between sampled points (emission regions) on each shell arriving at the observer to reduce numerical under sampling effects. Each emission region per shell is then attributed to a particular time bin and the luminosities in each bin are summed to create the final, composite light curve.

\subsection{Emission}
\label{subsec:Emission}

The energy of each emission region is calculated using 

\begin{equation}
    E_{i, \theta} = \eta_{\rm rad} \Gamma_{i, \theta} M_{i, \theta} c^2
    \label{eqn: region energy}
\end{equation}

\noindent where $M_{i, \theta}$ is the mass of each emission region and $\eta_{\rm rad}$ is the radiative efficiency which we set to 30\%, within the ranges derived in \cite{openingangles} and \cite{radiative_efficiencies1} and in agreement with that derived for GRB 190114C in \cite{radiative_efficiences2}. 

This energy is then distributed over a spectrum across an energy range of 15 to 350 keV \citep[chosen to match that of the \emph{Swift} Burst Alert Telescope BAT;][]{Swift_BAT}. 
We assume the Band function \citep{BandModel} as a standard GRB spectrum template. 
This is approximated by a sharply broken power-law with photon indices of $\alpha = 1$ and $\beta = 2.2$
either side of a peak energy, $E_p$.

To calculate $E_p$ for a given emission region we use the Amati relation \citep{Amati}, an empirically determined correlation between isotropic equivalent energy ($E_{\rm iso}$) and $E_p$. We apply a re-normalised version of the correlation determined for EE GRBs from \cite{EE_amatirelation} to each emission region per shell: 

 \begin{equation}
     E_{p, i, \theta} = 10^{b+2} (E_{i, \theta} / 10^{51})^a x,
     \label{eqn:Amati relation}
 \end{equation}
where $E_{p, i, \theta}$ is the peak energy of the spectrum applied at each emission region, the indices $a$ and $b$ have the values 0.38 and 0.82 respectively \citep{EE_amatirelation} and $x$ is a normalisation constant. 
Our normalisation constant accounts for the far lower energies associated with individual emission elements when compared to the time- and spatially-integrated ones typically associated with the Amati relation. We set $x$ = 200 in order to reproduce the Amati relation in the observed signal (EATS spectra) \citep{Amati, EE_amatirelation}.
 



This formalism calculates an energy release from mass and Lorentz factor without assuming a particular emission process, allowing our results to be generalised to the full suite of prompt emission models \citep[e.g.][]{internalcollisionsmodel, magneticreconnectionmodel, photosphericmodel, ICMARTmodel}.

Our model setup is axisymmetric, simulating a GRB viewed perfectly down the jet axis. In reality an on-axis observation is generally defined as a viewing angle anywhere within the core of the jet \citep{offaxis, insidelightcone, Ryan_structuredjet_closurerelations, structuredjets_review, coreviewingangles} and an off-axis viewing angle outside the core. However, this is beyond the scope of this paper. 
The effects of jet viewing angle will be explored in future works.

\subsection{Opacity}
\label{subsec:Opacity}

We calculate the optical depth to the photons emitted using the formalisms defined in \cite{LithwickandSari_Opacity}. A photon received by the observer must have an optical depth < 1. Optical depth decreases with increasing Lorentz factor. Therefore, from the energy of the received photon, a lower limit on Lorentz factor (the minimum value that allows the photon to escape) can be calculated. 

There are three sources of opacity: pair production from emitted photons (Limit A), photon scattering off the produced pairs (Limit B) and scattering off electrons associated with the baryons in the jet (Limit C). 

\subsubsection{Limit A: Opacity from photon annihilation}

A photon with energy $E'$ (in the co-moving frame) can only annihilate photons with energies larger than $\frac{m_e c^2}{E'}$. The maximum photon energy $E_{\rm max}$ therefore sets a minimum photon energy required to produce a pair through annihilation:

\begin{equation}
    E_{\rm max, an} = \frac{(\Gamma m_e c^2)^2}{E_{\rm max}}
    \label{eqn:threshold}
\end{equation}
where $m_e$ is the electron mass. 

We are interested in the local opacities at each emission region. Therefore in order to determine the number of photons above $E_{\rm max, an}$, ($N_{>E_{\rm max,an}}$), we need the local photon spectrum. As mentioned earlier, we approximate a Band function spectrum using a sharply broken power law. We apply this spectrum across an energy range $E_{\rm min}$ to $E_{\rm max}$, setting $E_{\rm min}$ to 1\,keV as a lower limit below which we expect negligible energy contribution from the spectrum and set $E_{\rm max}$ as 5$E_p$. This relationship between $E_p$ and $E_{\rm max}$  ensures $E_{\rm max}$ is tied to but significantly exceeds the characteristic photon energy of the spectrum. 

The optical depth 
is then

\begin{equation}
    \tau = \frac{11/180 \sigma_T N_{>E_{\rm max, an}}}{\pi \theta^2 R^2},
    \label{eqn:tau}
\end{equation}
where 11/180 is an approximation for the average cross-section \citep{Svensson_11/180approx}, $\pi \theta^2$ is the solid angle of the emission region and $R$ is the emission radius ($\pi \theta^2 R^2$ is therefore the physical size of the emitting region).

By demanding that photons with energy $E_{\rm max}$ have an optical depth < 1 we can derive a lower limit on the Lorentz factor of

\begin{equation}
    \Gamma_{{\rm min}, A} = \hat{\tau}^{1/(2\beta + 2)} \frac{E_{\rm max}}{m_e c^2}^{(\beta - 1)/(2\beta +2)},
    \label{eqn:limA}
\end{equation}

\noindent where $\beta$ is the photon index and $\hat{\tau}$ is a dimensionless quantity defined as 

\begin{equation}
    \hat{\tau} = \frac{\tau \Gamma^{2\beta + 2}}{(E_{\rm max}/m_ec^2)^{\beta - 1} }.
    \label{eqn:tau prime}
\end{equation}

This is adapted from Equation 4 in \citet{LithwickandSari_Opacity}, where it represents the optical depth for a photon of electron rest mass energy and a Lorentz factor of order unity.


\subsubsection{Limit B: Compton scattering from produced pairs}

Pairs produced by the annihilation of higher energy photons contribute to the electron and positron field that leads to limit B. 
This source of opacity comes from Compton scattering of photons off the electron and positron field. This applies to photons of all energies. 

The number of pair producing photons can be approximated by the number of photons that can self-annihilate (annihilate with a photon of the same energy), $N_{>\rm self, an}$. In order to account for the particles produced by previous shells, we sum the annihilated photons from previous shells as well as the current shell. 

Above a certain number of pair producing photons the emission is optically thick to pair scattering. \cite{LithwickandSari_Opacity} term this number $N_{>\rm thick}$. This is the number of photons with an energy above $E_{\rm thick}$, where $E_{\rm thick}$ is the photon energy for which the optical depth is exactly one. 

For the emission to be optically thin to this scattering opacity $N_{>\rm self, an}$must be less than $N_{>\rm thick}$. Defining this in terms of energy, $E_{\rm self, an}$ < $E_{\rm thick}$ and a limit on Lorentz factor can be derived 

\begin{equation}
    \Gamma_{{\rm min},B} = \hat{\tau}^{\frac{1}{\beta + 3}}
\end{equation}
where $\hat{\tau}$ is defined in Equation~\ref{eqn:tau prime}. 

Emission regions with $\Gamma < \Gamma_{{\rm min},B}$ are optically thick to the emitted photons and therefore do not contribute to the final light curve. 

\subsubsection{Limit C: Scattering off electrons associated with baryons}

\cite{LithwickandSari_Opacity} determine a lower limit on the number of baryons by assuming the total energy in photons is less than the kinetic energy of the baryons: $E_{\rm min} N_{>\rm min}$ < $N_{\rm baryons}\Gamma  m_p c^2$. The optical depth can then be calculated using Equation~\ref{eqn:tau} and a lower limit on Lorentz factor obtained from the condition that $\tau$ < 1. 

Limit C can be computed in conjunction with limit B through an additional pre-factor

\begin{equation}
    \Gamma_{{\rm min}, B C} = [(\hat{\tau}^{\frac{1}{\beta + 3}})^{\beta - 2} \frac{E_{\rm min}}{m_p c^2}]^{1/5} \Gamma_{{\rm min},B}.
\end{equation}
However, we find, in agreement with \cite{LithwickandSari_Opacity}, that limit C is much less significant, with the additional prefactor of order unity.

\begin{figure}
	\includegraphics[width=\columnwidth]{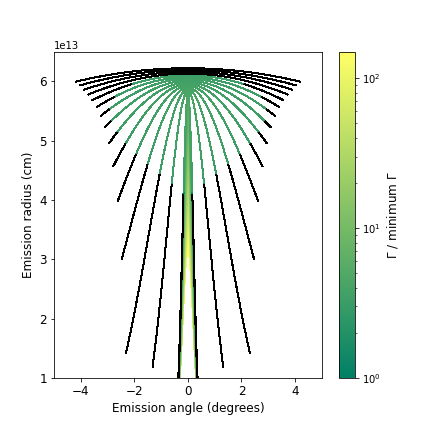}
    \caption{Diagram of individual shells within the jet showing the trapped emission (optical depth > 1) in black. Each line is an individual shell with emission regions plotted as angle from central axis against emission radius. The colour bar denotes the Lorentz factor across the shells as a fraction of the minimum Lorentz factor from the most stringent opacity limit. The brighter regions with $\Gamma / \Gamma_{\rm min}$ > 1 are where the optical depth is smaller than one and hence the emission escapes. The darker regions with $\Gamma / \Gamma_{\rm min}$ < 1 are where the optical depth is larger than one, hence the emission is trapped and does not reach the observer. }
    \label{fig:opacity_jetshape}
\end{figure}

Figure~\ref{fig:opacity_jetshape} shows which parts of the jet have optical depth less than one and hence emission can escape. As in Figure~\ref{fig:JetShape} a small subset of individual shells are plotted as angle from central axis ($\theta$) against emission radius. The colour bar now displays Lorentz factor as a fraction of the limiting Lorentz factor due to opacity constraints, where the most stringent limit has been applied at each point. Emission from lighter regions (with $\Gamma/\Gamma_{\rm min}$ > 1) escapes whereas emission from the black regions (with $\Gamma/\Gamma_{\rm min}$ $\leq$ 1) is trapped. 

As the Lorentz factors in the Gaussian jet drop, the emission from the wings is lost to opacity. The wings of the initial few shells have high enough Lorentz factors to escape however, the wings of the last shells launched as the Gaussian jet, before and during jet expansion, are truncated. Emission from the lowest Lorentz factor shells (launched last) also does not escape.

\section{Light Curves}
\label{sec:lightcurves}

\begin{figure}
	\includegraphics[width=\columnwidth]{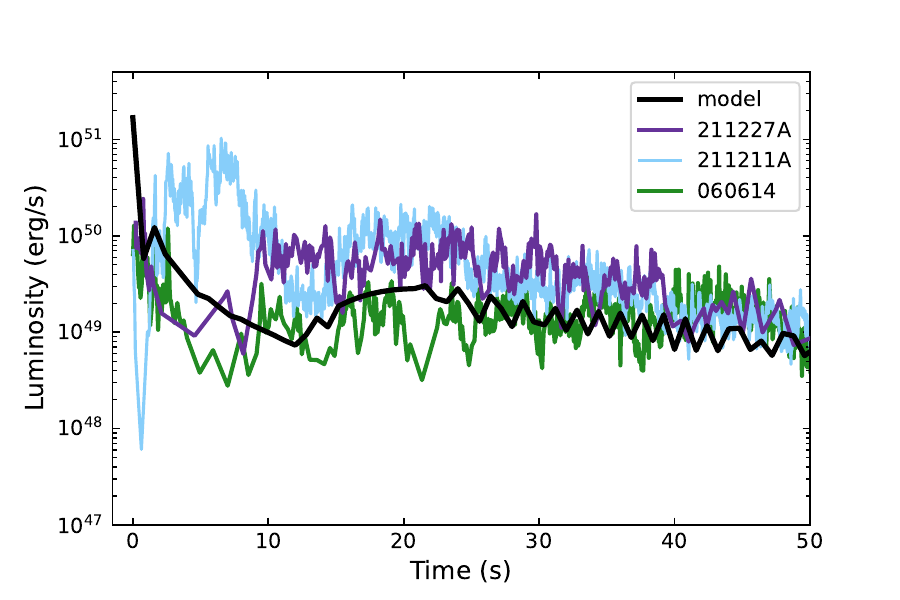}
    \caption{Data of three representative long merger GRBs detected by \emph{Swift}: 211211A (blue), 211227A (purple) and 060614 (green). Plotted as well is our canonical model light curve in black (created using the parameters in Table \ref{tab:canonical params}).}
    \label{fig:data+model}
\end{figure}

Here we present light curves produced from our model. Our light curves are integrated across a bandpass of 15 - 350\,keV \citep[matching that of the \emph{Swift} Burst Alert Telescope BAT;][]{Swift_BAT}.  

\begin{table}
	\centering
	\caption{Free parameters in the model with the values used to create our canonical light curve.}
	\label{tab:canonical params}
	\begin{tabular}{ccc} 
        \hline 
        N& 1000 & number of shells\\
		\hline
		$\Gamma_0$ & 300 & maximum Lorentz factor\\
		\hline
		$\kappa$ (s) & 0.1 & turnover time\\
        \hline 
        R (cm) & 10$^{13.5}$ - 10$^{15}$ & emission radii \\
        \hline
        $\theta$ ($^\circ$) & 2-5 & jet size\\
        \hline 
        $\theta_c$ ($\theta$) & 0.1 - 5 & core size\\
        \hline
        $\Gamma_t$ & 9 & trigger Lorentz factor\\
        \hline
        s (shells) & 20 & smoothness\\
	\end{tabular}
\end{table}

Figure~\ref{fig:data+model} shows our canonical model light curve superimposed over the data of three representative EE GRBs detected by \emph{Swift} \citep{Swift}: 211211A, 211227A and 060614. The light curves of the representative GRBs are taken from the UK \emph{Swift} Science Data Centre \citep{Evans07,Evans09} and use the evolving photon index to extrapolate the reported flux densities across the BAT bandpass. The model parameters are quoted in Table~\ref{tab:canonical params} and are tuned to match the comparison sample while being consistent with the literature.

The fiducial model successfully reproduces the broad features of EE bursts showing an initial IPC-like spike followed by a more plateaued EE-like section. The primary aim of the model is to produce the EE signature. The onset time of the EE rebrightening hump ($\sim$ 10\,s) closely matches GRB 211211A in particular, with a decay slope similar to both 211211A and 211227A. The luminosity of the later EE section is consistent with all three observed bursts. The IPC luminosity is comparable to GRB 211211A while slightly higher than GRBs 060614 and 211227A. We do not attempt to reproduce a long, complex IPC like that of GRB 211211A, focusing instead of the later EE signal. The IPC is too spiky to be produced by a smooth blending of shells with constantly decreasing Lorentz factors and is therefore beyond the scope of this model. 

The IPC is produced by shells within a narrow cored, Gaussian structured jet. The narrow core size of 0.1$\theta$ creates a steeply structured jet where the Lorentz factors drop off rapidly towards the wings. The core of the jet is therefore the dominant contributor to the early emission. Initially the Lorentz factors in the core are of order 100, with a maximum Lorentz factor of 300 from the first launched shell \citep[within the range of Lorentz factors derived from compactness constraints and GRB fitting;][]{LorentzFactors, fireballmodel, LithwickandSari_Opacity, gammaspread}. These produce the initial bright spike. 

Higher Lorentz factors create more strongly beamed emission corresponding to smaller beaming cones. The first shells therefore produce less HLE as the observable angle is more restricted. As the Lorentz factors decrease with each subsequent shell, emission from further along the wings becomes observable as the lower Lorentz factors at the wings, due to the lateral structure, have larger beaming cones. However, as mentioned in the previous section, Figure~\ref{fig:opacity_jetshape} shows the wings of these shells are below the minimum Lorentz factor required for emission to escape. Therefore, the pulse from each shell is short, with minimal contribution from the wings. The steep drop off after the initial spike hence reflects the exponential launch distribution of Lorentz factor as the main contributions remain from the core of the jet. 

The EE is produced by a wider, tophat-like structured jet. We approximate a tophat structure by setting the core size parameter to $\theta_c = 5\theta^\circ$. This effectively sets the Lorentz factors at the edges of the jet to be comparable to those at the centre. 
The entire $5^\circ$ jet is then comparable to the core size of 6$^\circ$ used by \cite{Gavin_170817} to model GRB 170817A. 

The onset of the jet expansion is controlled by the trigger Lorentz factor, which we set as $\Gamma_t = 9$. The Lorentz factors in the wider jet that produces the extended emission are therefore $\leq$ 9. Lower Lorentz factors allow for more HLE per shell and therefore each pulse is extended in time. The tophat structure also increases the Lorentz factors at wider angles relative to the core, making the contributions from HLE more pronounced. The superposition of these longer, more energetic pulses creates the plateaued nature of the EE section of the light curve, with the increase in their overlap smoothing out each individual onset. The wider the core size, the more energy is distributed to the wings, and hence the brighter and more plateaued the observed emission (see Section~\ref{sec:core size}). 

The jet expansion can also be considered in terms of the time the wider shells are launched using Equation~\ref{eqn:trigger time}. When the Lorentz factor at the head of the next shell is $\leq \Gamma_t$ (or equivalently $t \geq t_{\rm trigger}$), the widening core size and expanding jet width effects are triggered. The onset time of the EE signature is mainly controlled by the combination of this parameter and $\kappa$, the steepness of the Lorentz factor and mass launch distributions. Their values are chosen to best match the EE onset times of the bursts in Figure~\ref{fig:data+model} (in conjunction with the maximum Lorentz factor). The rate of jet expansion is controlled by the smoothness parameter ($s$), a numerical parameter detailing the number of shells over which the jet expansion takes place. This prevents a sudden and unphysical jump in the jet and core width. The transition from a Gaussian to a tophat structure takes place over twenty shells, chosen to remove an artificial sharp rise without inducing additional numerical expense.

A tophat jet (at the same central Lorentz factor) as a Gaussian jet will produce brighter emission as the Lorentz factor remains high across the entire jet width. However, emission from the wider angles will still reach the observer with less energy because not all the emission is directly beamed into the observer's line of sight. The wider the jet therefore, the lower luminosity emission from the wings produces. The jet widths were therefore chosen to balance the interplay between bright emission allowing the EE re-brightening to exist while still requiring the jet to expand in synchronicity with the widening core size. Our initial narrow jet of $2^\circ$ is consistent with previously modelled, narrow GRB jets \citep{narrow_jet_core, bubbles, openingangle_GRB061201, openingangle_GRB160325A} while our wider jet of $5^\circ$ falls within ranges derived from jet breaks \citep{jetbreak_openingangles, jetbreak_openingangles2, openingangles}.

We tested the scenario of a single emission radius for all shells and a small range of radii where the later shells emit further from the central engine (see Section~\ref{sec: Emission radii}). The small range of radii accentuates the EE signature. The later shells have slightly further to travel before they emit and hence their emission is received with more delay with respect to the earlier shells that create the IPC. This results in a slightly more emphasised EE onset. We therefore use a fiducial emission radius range covering the tested emission radii for which the EE signature is apparent. Our values are consistent with emission radii predicted by internal collision shell models \citep[$\sim 10^{13} - 10^{14}$\,cm;][]{internalcollisionsmodel, ICMARTmodel, Lazzati_jet+cocoon_model}.

The number of shells affects the spikiness of the light curve. Launching more shells over the same period of time increases the overlap between the individual shell pulses and therefore smooths out variability in the light curve. A small number of shells can therefore induce numerical variability. However, there is also a physical element to this variability. Observed light curves exhibit spiky behaviour with small minimum variability timescales (MVTs) that are often assumed to be related to the activity of the central engine \citep{internalcollisionsmodel, MVT_internalshocks, MVT}. We choose 1000 shells to balance between these physical and numerical effects.

\subsection{Parameter Exploration}
\label{sec:parameter exploration}

This section explores the parameter space to identify how specific parameters govern the morphology of the light curves. In all the following figures, the black line is our canonical light curve (using the parameters in Table~\ref{tab:canonical params}), shown for comparison.

\subsubsection{Maximum Lorentz Factor}
\label{sec: max Lorentz factor}

\begin{figure}
    \includegraphics[width=\columnwidth]{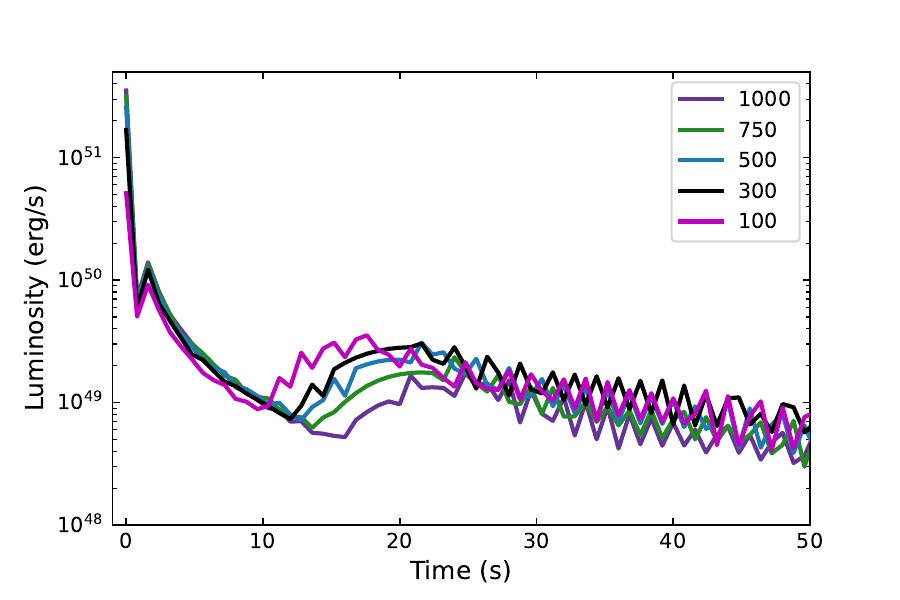}
    \caption{Exploration of the impact of the maximum Lorentz factor of the emitting shells. The black line is our canonical light curve (the same as that compared to the data in Figure \ref{fig:data+model}). The value of every other parameter is taken from Table \ref{tab:canonical params}.}
    \label{fig:max gammas}
\end{figure}

Figure~\ref{fig:max gammas} shows an exploration of the maximum Lorentz factor of the shells ($\Gamma_0$ in Equation~(\ref{eqn:lorentz factor distribution})). The Lorentz factor affects the brightness of the initial peak, a larger Lorentz factor means more energy is distributed to the shells. 

In contrast to the initial peak, the inverse effect occurs in the EE part of the light curve where larger Lorentz factors result in a fainter received signal. Larger Lorentz factors do, of course, inherently produce brighter emission. This is however offset by a smaller Doppler boost (Doppler boosting is inversely proportional to Lorentz factor (Equation~\ref{eqn:Doppler})) and less HLE received by the observer due to relativistic beaming effects. Larger Lorentz factors at the wings of the shells lead to narrower beaming cones (since the width of the cone is inversely proportional to the Lorentz factor). Sight lines that intersect the observer are therefore restricted to a narrower portion of the jet, reducing the emitting area from which photons are received. At this point in the light curve the HLE contributes more significantly to the received emission than earlier (during the IPC) and therefore the reduction in this emission has a stronger effect, reducing the overall received flux.

The timing of the EE re-brightening also changes with Lorentz factor. This is a consequence of the model set-up rather than a real effect in this case. The widening of the jet and core that produces the re-brightening is activated when a pre-defined threshold is met for the Lorentz factor (for example, in the canonical light curve, when the subsequent shell has a Lorentz factor $\leq$ 9 the widening is triggered). This trigger parameter was left unchanged and hence the larger starting Lorentz factors take longer to reach that trigger, pushing the re-brightening later in time. 

The jet expansion causes a comparable luminosity rise across all Lorentz factor values. It can therefore be inferred that this feature is not dependent on the absolute values of the Lorentz factors.


\subsubsection{Trigger Lorentz Factor}
\label{sec: Trigger gamma}

\begin{figure}
    \includegraphics[width=\columnwidth]{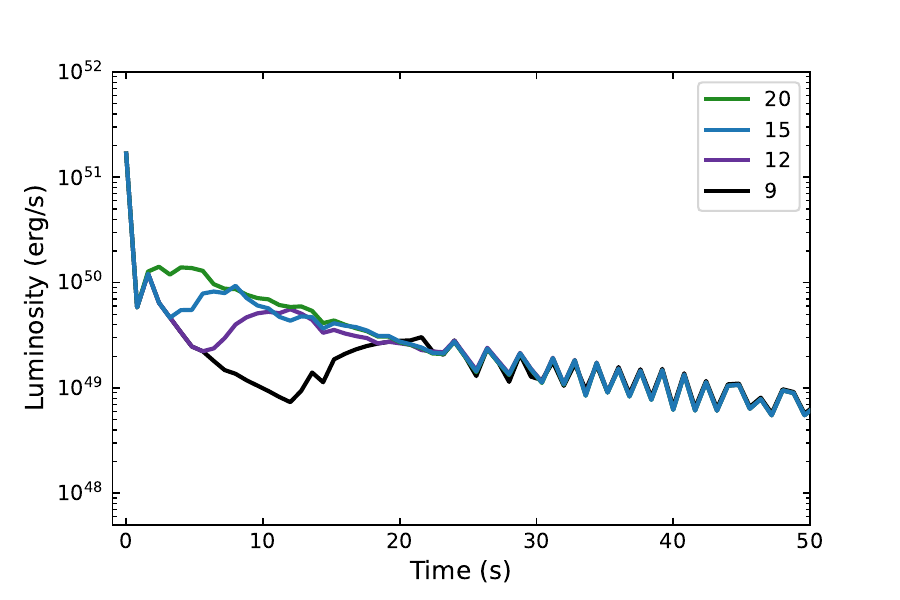}
    \caption{Parameter exploration of our trigger Lorentz factor ($\Gamma_t$). This parameter controls when the jet expansion occurs and is therefore the maximum Lorentz factor in the wider jet. The different lines are different values of $\Gamma_t$ and the black line is our canonical light curve (the same as that compared to the data in Figure \ref{fig:data+model}). The value of every other parameter is taken from Table \ref{tab:canonical params}.}
    \label{fig:trigger gamma}
\end{figure}

From Figure~\ref{fig:trigger gamma} it can be seen that in order to produce an EE-like re-brightening the Lorentz factors in the wider jet need to be a maximum of 12 (purple line in the figure). If the Lorentz factors are much higher the EE signal becomes sub-dominant compared to the early shells and its signature is lost in the light curves.

This parameter sets the time of the jet expansion (see Equation~\ref{eqn:trigger time}) and hence for lower values the EE rise is seen later. The value of $\Gamma_t$ is inherently degenerate with the maximum Lorentz factor. It is therefore the ratio between $\Gamma_0$ and $\Gamma_t$ that determines the EE timing. However, as mentioned in the previous section (Section~\ref{sec: max Lorentz factor}) larger Lorentz factors produce less HLE. In the EE part of the light curve the HLE is most dominant. This therefore creates a maximum limit on the Lorentz factors in the wider jet at $\sim$ 12. We find the ratio of 300/9 best creates an EE signal to match the data.

\subsubsection{Turnover Time}
\label{sec: Turnover Time}

\begin{figure}
    \includegraphics[width=\columnwidth]{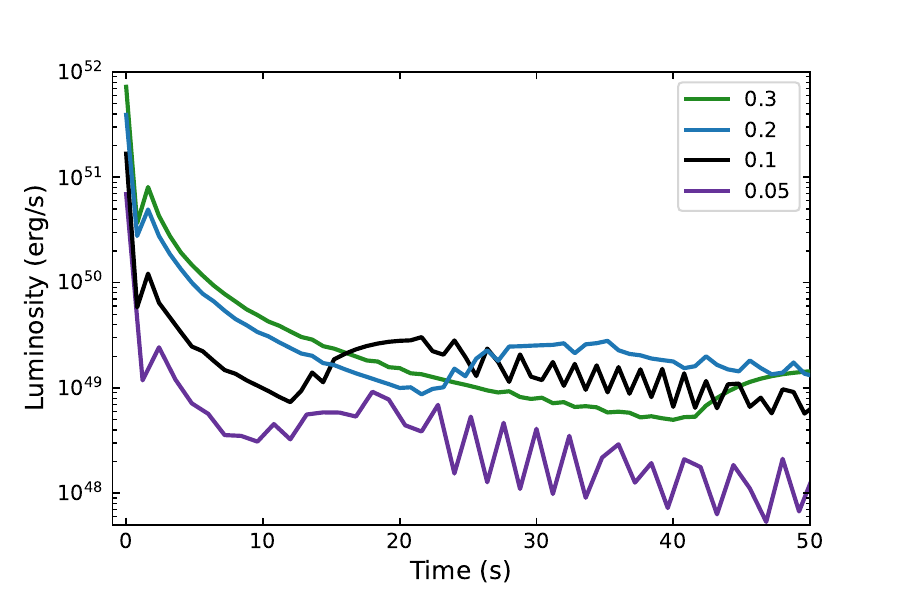}
    \caption{Exploration of the impact of the turnover time of the exponential launch functions. The black line is our canonical light curve (the same as that compared to the data in Figure \ref{fig:data+model}). The value of every other parameter is taken from Table \ref{tab:canonical params}.}
    \label{fig:tau}
\end{figure}

The turnover time, $\kappa$, controls the launch distributions of Lorentz factor and mass. It represents the time the distributions turn over and hence, the smaller its value, the faster these parameters evolve. The energy in the shells therefore decreases faster. This can be seen in Figure~\ref{fig:tau} where the smallest value (purple line) produces the weakest emission. At these Lorentz factors the EE rise is barely distinguishable since the energy in the later, wider shells is reduced. 

Larger values of $\kappa$ mean the trigger Lorentz factor criterion for jet expansion, $\Gamma_t$, (discussed above) is achieved by a later shell and hence the EE rise appears at later times. For $\kappa \geq$ 0.3, the rise is not present within the range seen in our observed bursts.



\subsubsection{Emission Radii}
\label{sec: Emission radii}

\begin{figure}
    \centering
    \includegraphics[width=\columnwidth]{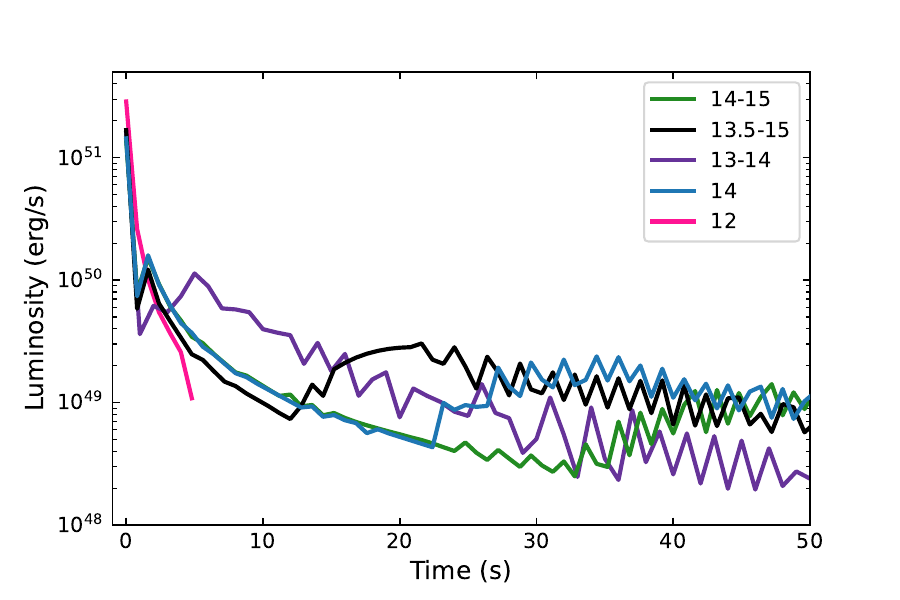}
    \caption{Exploration of different emission radii (all other parameters are taken from Table \ref{tab:canonical params}). The black line is our canonical light curve. A radius range of $10^{12}$ $<$ R $>$ $10^{15}$cm produces a light curve with an initial prompt spike plus some longer lasting emission. Below this range the emission does not last long enough to produce EE.}
    \label{fig:radius range}
\end{figure}

Figure~\ref{fig:radius range} shows our canonical light curve (in black) compared to different emission radii. 

The model is set up such that the shells emit when they reach their pre-defined emission radius. The emission radius is therefore partly responsible for the delay time between subsequent shells emitting (as well as the relative Lorentz factor of the shells). 

A smaller radius produces a brighter initial peak because the delay time between subsequent pulses arriving is smaller. The emission is therefore spread over fewer time bins, creating a brighter summed luminosity. Our model does not include any deceleration of the shells so this effect is due solely to the reduced travel distance.  

However, because the delay times are shorter, the later, lower Lorentz factor, wider shells that produce the EE arrive earlier, truncating the overall duration of the light curve. In order to produce the longer durations that characterise EE, the emission radius needs to be larger than $10^{12}$cm. 


The range of emission radii for which the model can produce extended emission-like light curves, within the timespan seen in our observed bursts, is $10^{12}$ $\leq$ R $\leq$ $10^{15}$cm. This range satisfies predicted prompt emission radii values from several different models including internal shock models where the radius is expected to be between $10^{13}$cm and $10^{14}$cm \citep{internalcollisionsmodel, ICMARTmodel, Lazzati_jet+cocoon_model} and the ICMART model where larger emission radii are required ($10^{15}$ - $10^{16}$cm) \citep{ICMARTmodel}.

This range of radii sits below an upper limit set by the deceleration radius of the first shell calculated using \cite{Nava2013_Rdec}:



\begin{subequations}
\begin{align}
    A = \frac{1377}{4096\pi} \frac{ E} {n m_p c^5} \\
    R_{\rm dec} = \frac{4}{3} c (1+z) (A^{-0.5} \Gamma_0)^{-2/3}
\end{align}
\end{subequations}

where $E$ is the kinetic energy in the burst (approximated as $E_{\rm iso}$), $n$ is the nucleon number density \citep[taken as a standard value of $1\,cm^{-3}$;][]{n_afterglowvalue}, $m_p$ is the proton mass, $z$ is the redshift \citep[where we have used the redshift of GRB 211211A, $z = 0.0763$;][]{211211A_kilonova_Jillian}, $\Gamma_0$ is the Lorentz factor at the head of the leading shell and $R_{\rm dec}$ is the deceleration radius of the leading shell.

\subsubsection{Core Size}
\label{sec:core size}

\begin{figure}
    \centering
    \includegraphics[width=\columnwidth]{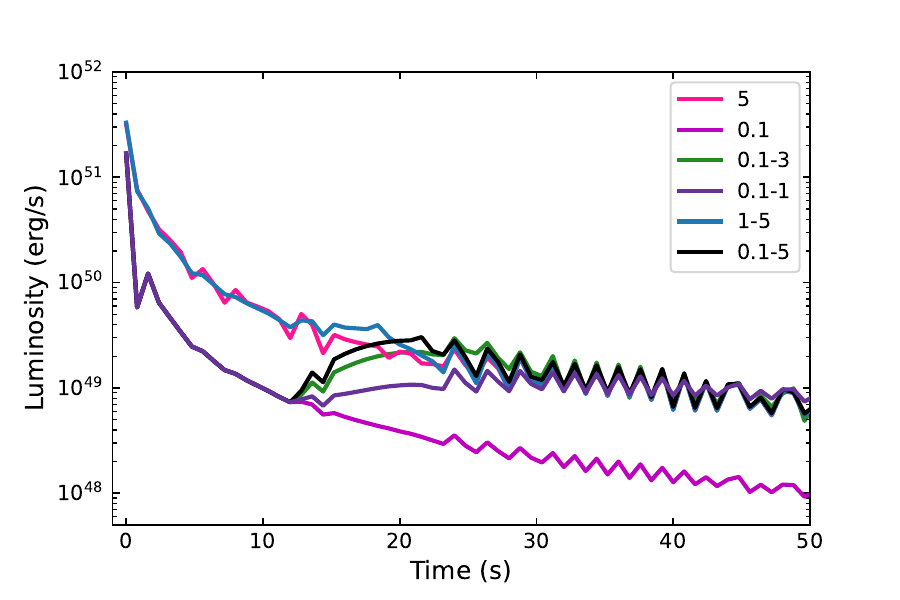}
    \caption{Testing the lateral structure of the jet by adjusting the core size. All other parameters are taken from Table~\ref{tab:canonical params}. These light curves are computed with a widening jet expanding from 2 to 5$^\circ$ (matching the jet width in Table \ref{tab:canonical params} for the canonical light curve). The pink and magenta light curves assume a fixed jet structure (tophat and Gaussian respectively) that never changes: the jet widens but the relative lateral structure remains constant (the relative core size remains the same, growing only in proportion with the jet width). The other light curves have an accompanying lateral structure change as the jet expands (parameterized as core size). The black line corresponds to our canonical light curve created from the values in Table~\ref{tab:canonical params}.}
    \label{fig:coresize}
\end{figure}


In our canonical light curve we have a widening jet accompanied by a change in the lateral structure, from a Gaussian to tophat jet, mimicking a jet+cocoon or two-jet-like structure \citep{Jin_2componentoptions, vv_2jet, 170817_jet+cocoon_model, jet_cocoon_model_salafia}. The upper limit of the core size (5) approximates a tophat jet with minimal lateral structure and a near-constant Lorentz factor with angle as the Lorentz factors at the edge of the jet are comparable to those at the centre. The lower limit (0.1) creates a narrow cored, Gaussian structured jet.

Figure~\ref{fig:coresize} explores the effects of changing the lateral structure. The pink and magenta light curves have their core sizes fixed to the final and initial values in our canonical light curve respectively, enforcing a constant lateral structure for their duration. The pink light curve is a constant tophat jet and the magenta a constant Gaussian. Their lack of re-brightening suggests that a changing lateral structure is essential to produce EE. Furthermore, by contrasting the models with changing jet structure (remaining light curves), it is clear that the magnitude of the structure change is an important factor, where a larger ratio between the initial and final core sizes results in a more pronounced EE effect. A narrower initial core (steeper distribution) means the IPC drops off more steeply, increasing the luminosity contrast with the EE component.

The blue light curve incorporates a structure change from a Gaussian to a tophat jet, however, the core size of the initial jet is larger meaning the Gaussian profile is less steep. This leads to larger Lorentz factors in the wings of the initial shells which then provide a more dominant contribution to the later light curve masking the EE re-brightening feature. This light curve demonstrates that a narrow cored initial jet is necessary to create the EE feature.

\subsubsection{Opening Angle}
\label{sec:opening angle}

\begin{figure}
    \centering
    \includegraphics[width=\columnwidth]{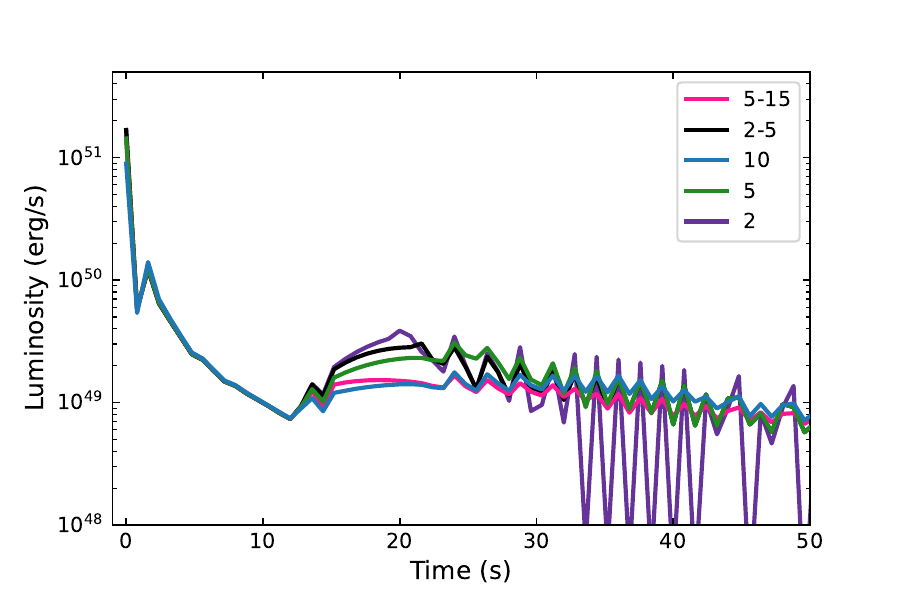}
    \caption{Exploration of the opening angles of the jet. All other parameters are taken from Table~\ref{tab:canonical params}. The values in the legend refer to the initial and final opening angles of the jet before and after the widening. The black line is the canonical light curve created with the parameters in Table~\ref{tab:canonical params}.}
    \label{fig:opening angle}
\end{figure}

GRB jets typically have opening angles of a few degrees \citep{openingangle_GRB061201, openingangles, openingangle_GRB160325A}. Here, we experiment with different opening angles and investigate how they impact the morphology of the light curves the model produces. 

The light curves in Figure~\ref{fig:opening angle} show a decrease in luminosity with opening angle. A wider jet increases the emitting area. Each emitting element is now travelling at a more oblique angle with respect to the observer. Doppler beaming (Equation~\ref{eqn:Doppler}) has an angular dependence where emission from wider angles is less boosted since a smaller fraction of the light emitted is directed towards the observer. The observer therefore receives a lower flux from a wider jet.


\section{Discussion}
\label{sec:discussion}

There is large variation in observed EE GRB light curves and they can therefore be hard to identify. There does however appear to be a general, recognisable structure of a main peak followed by a temporally extended, spectrally softer, smoother hump of emission lasting up to 10s of seconds with the two sections separated by a quiescent gap \citep{EE_GRBs, Dichiara2021, EE_MVT}. 

Our model light curves are able to recreate this general structure, displaying a short peak IPC $\leq$ 2\,s followed by a rising, smoother hump with a dip in emission between the two sections. The HLE is most dominant in the later part of the light curve, contributing to the EE signal, amplified by the jet expansion from a narrow-cored Gaussian to a wider cored, more tophat-like jet. 

\cite{EE_MVT} show that in their sample the EE has a larger MVT than the prompt emission, contributing to a smoother signal. The smoothness of the light curves produced by our model is of course partially dependent on the numerical sampling chosen. The lower Lorentz factors of the shells producing the later EE signal however mean the arrival times between the shells are increased. This would lead to an increased MVT at later times. Indeed the numerical undersampling effects are more visible in the later parts of the light curves as the bin size becomes smaller in comparison with the emission arrival time differences. 

Our distribution of Lorentz factors within the jet, that leads to the lower Lorentz factors that create the EE, is consistent with the stratified outflow used in \cite{X-rayplateaus_radialgamma}. They model X-ray plateaus, commonly seen in extended emission bursts \citep{X-rayplateau_lateprompt, Ben_expaccretion} also from the accretion energy reservoir, as we do for the earlier extended emission signal, tying the prompt, X-ray plateau and early afterglow to the same outflow. Jets with a distribution of ejecta Lorentz factors have also been invoked as energy injection for afterglow modelling \citep{Rees1998_stratifiedjet, Kumar2000, Sari2000_stratifiedjet}. In particular, refreshed shock models where more energetic but lower Lorentz factor shells are ejected later have been proposed to explain afterglow plateaus and flares \citep{GRB160821B_refreshedshock, GRB210726A_radioflare, GRB231117A_refreshedshock}. Our model launches higher Lorentz factor shells first as part of the narrow jet, followed by lower Lorentz factor shells as part of the wider jet. 

\subsection{Jet Expansion}

The changing lateral structure (core size) is the most influential parameter controlling the EE onset rise. Several options for a changing lateral structure exist in the literature. The initial fast and narrow jet could clear a path for the wider, slower jet. With less material to plough through the wider jet retains more energy. This has been proposed by \cite{EEfromfallback} who invoke additional fallback material to power the EE that launches a wider jet in the space evacuated by the narrower jet responsible for the early emission.  

In \cite{Rosswog2003} a thick disk was considered to collimate the jet and \cite{Jin_2componentoptions} suggested a potential scenario where the disk became unstable as the jet was launched creating a more baryon-loaded, less collimated second jet component to power the later afterglow emission. Additionally \cite{Jin_2componentoptions} suggest a 2-component jet instead, following the prescription of \cite{neutron_proton_2componentjet} where the outflow has a high neutron to proton ratio resulting in a lower Lorentz factor, wider neutron dominated jet while the protons are able to be accelerated to higher Lorentz factors and are hence collimated into a narrower angle by the electromagnetic force. Alternatively, \cite{vv_2jet} created a model to explain EE bursts involving a $\sim$ 6$^\circ$ jet powered by neutrino annihilation followed by a narrower electromagnetic jet powered by the BZ mechanism \citep{BZpaper} to explain the hard spectra IPC and the softer EE respectively. \cite{2component_earlyafterglow} use a 2-component jet model consisting of a narrow jet of width 0.4$^\circ$ and a wider jet of 8$^\circ$ to explain the early afterglow behaviour of GRB 080319B.

There is also the potential for an interaction between the jet and the kilonova ejecta. \cite{Nicholl_kilonovamodel} suggest that kilonova models might require an interaction between the kilonova ejecta and a cocoon where the ejecta becomes shock heated. Such an interaction affects the dynamics of the jet and could contribute to the wider jet obtaining more energy. A jet plus cocoon style structure has been routinely applied to GRB 178017A \citep{170817_jet+cocoon_model, 170817_Gottliebcocoon, 170817_gaussianjetmodel_+baryonloading} where shock interactions between the jet and the merger ejecta deposit energy into the surrounding material creating the cocoon. 


\subsection{EE and IPC Fluence and Luminosity Comparisons}

Figure~\ref{fig:data+model} shows that the EE part of the light curve (around 10\,s) is fainter than the initial IPC spike by about two orders of magnitude. Here we investigate the range of EE to IPC ratios the model can produce and compare to observed data. We parameterize the ratio by computing an EE fluence over an IPC fluence (EE/IPC). Motivated by the typical maximum short GRB $T_{90}$, we define the IPC for this calculation as less than 2\,s. We define EE as between 10 - 50\,s based on the time the rise appears in our light curves. 

A wide range of observed EE to IPC fluence ratios have been reported. \cite{Perley2009} find values covering a range of EE/IPC = 0.3 - 30, consistent with the range of 0.21 - 33.5 with a median value of 1.7 found by \cite{kaneko2015}. Our canonical light curve has a ratio of 0.4 with our largest ratio derived from the magenta light curve in Figure~\ref{fig:max gammas} (which has a maximum Lorentz factor of 100) having a value of 1.1. Our light curves are produced assuming an on-axis observer. If the EE/IPC ratios are a product of viewing angle, we would expect lower ratios from our on-axis model. An on-axis viewing angle enhances the IPC which is dominated by the on-axis emission. Our EE to IPC fluence and luminosity ratios are therefore their minimum values; off-axis viewing angles will have larger ratios as the EE is enhanced with larger Doppler boosts and the emission producing the IPC is not directly beamed to the observer. 

\cite{EE/IPC_intensity} considers the relative brightness of the IPC and EE in terms of intensity rather than fluence. They found that the EE is generally lower by a factor of $10^{-3}$ to $10^{-2}$. While, for our canonical light curve, our fluence ratios are towards the smaller end of the ratios found by \cite{Perley2009, kaneko2015}, our peak luminosities of the IPC and EE are at the top end of this range, with our EE onset luminosity $\sim$ $10^{-2}$ times the IPC peak. 

As well as the viewing angles effects mentioned above, smaller fluence ratios might, at least in part, arise from different prescriptions for assigning IPC and EE duration since neither \cite{Perley2009} nor \cite{kaneko2015} outline their exact methods for determining durations. Additionally, we compare our theoretical light curve to observed light curves without considering whether the entirety of the IPC would be captured, either through an off-axis viewing angle or emission outside the observing bandpass, which could potentially alter our IPC fluence. The jet expansion in our model gives us a tool to vary this EE/IPC ratio and allows us to create a range of possible ratios. 

\subsection{Short GRBs}
\label{subsec: Short GRBs}

The smallest EE/IPC ratio in our model (from $\Gamma_0$ = 1000), 0.1, corresponds to a decrease of 2.5 orders of magnitude in luminosity. Imposing a simple, lower sensitivity limit, it is possible for the EE to fall below, leaving only the IPC detected. The received signal would therefore appear to be a standard short GRB without extended emission. 

From the parameter exploration performed: a larger emission radius, smaller turnover time and wider jet width can all add to the reduction in luminosity of the EE section of the light curve. Within a certain parameter space, the EE emission could potentially sit below the observing bandpass, creating a short GRB like signal. Thereby a continuum from short to extended emission light curves can be produced from the model.

\subsection{Spectral Evolution}
\label{subsec: Spectral Evolution}



\begin{figure}
    \centering
    \includegraphics[width=\columnwidth]{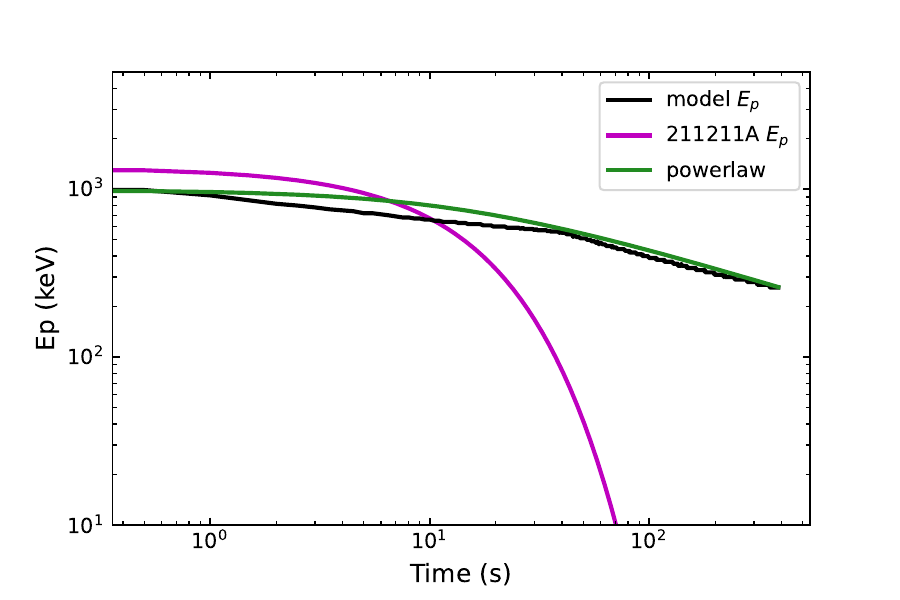}
    \caption{Evolution of $E_p$ from the outputted model spectra. The  The magenta line shows the evolution of the peak energy inferred from the data of GRB 211211A \citep{Ben211211A}. The green line shows a power-law decay with the same characteristic turnover time (14.4\,s) as the magenta line.}
    \label{fig:Ep}
\end{figure}

Long mergers typically show a hard to soft spectral evolution throughout their duration where the EE is spectrally softer than the IPC \citep{EE_GRBs, Troja_NSBH, vv_2jet, kaneko2015}. We investigate this by looking at the evolution of $E_p$ over time from our EATS spectrums, shown in Figure~\ref{fig:Ep}. The magenta line shows the evolution of $E_p$ measured in GRB 211211A: an exponential function with a characteristic turnover time of 14.4\,s \citep{Ben211211A}. The green line is a smoothly broken power law with the same characteristic turnover time. While the turnover time of the model $E_p$ (black line) matches the 211211A turnover time fairly well, the subsequent spectral evolution is not as steep, matching the power law much better than the exponential from 211211A. The model induces some spectral evolution but at a slower rate than that measured through observations. 

This may suggest some other physical conditions within the jet or its environment that are not currently accounted for in the model are changing and inducing the late-time evolution. For example, the particle number density or the magnetic field energy factor \citep{ChangingB_field, Ronchini_HLE}. The investigation of time-varying environment parameters is beyond the scope of the current model and is deferred to future work.

\begin{figure}
    \centering
    \includegraphics[width=\columnwidth]{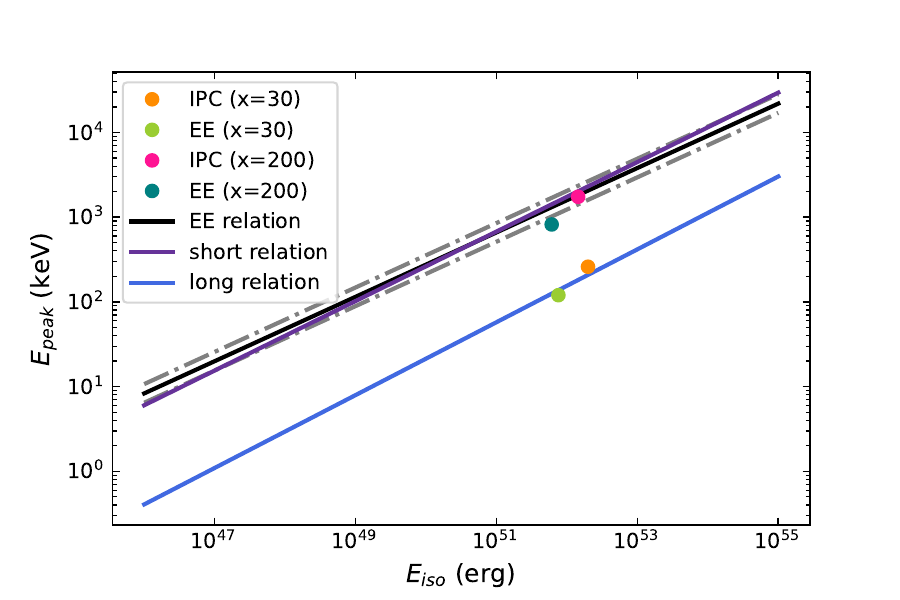}
    \caption{The Amati relations for long, short and EE bursts are shown in blue, purple and black respectively, taken from \citep{EE_amatirelation}. The positions of the IPC and EE from our canonical model light curve (computed with the parameters from Table~\ref{tab:canonical params}), for two different normalisations, $x$, of our $E_p$, are plotted alongside. A normalisation of 200 was used to create the light curves presented in Section~\ref{sec:lightcurves}. IPC points (pink and orange) include all emission up to 2\,s while the EE points (blue and green) contain the remaining emission from 2\,s onwards. The time integrated $E_{p}$ is approximated as the maximum within the time region considered. The time integrated $E_{\rm iso}$ is determined by summing the $E_{\rm iso}$ at each EATS (calculated by multiplying the luminosity across the EATS by the time resolution).}
    \label{fig:Amati}
\end{figure}

We plot the position of the IPC and EE from the EATS spectra on the Amati relation in Figure~\ref{fig:Amati}, for two different $E_p$ normalisations, where we include the Amati relations for long (in blue) and short (in purple) GRBs as well as EE bursts (in black) for comparison. The IPC points (in pink and orange) include emission up to 2\,s and the EE points (in blue and green) all emission after 2\,s. We approximate a time-integrated value of $E_{p}$ by taking the maximum value across the EATS spectra within the time ranges. $E_{\rm iso}$ is determined by summing the product of luminosity and time resolution across the EATSs.  

The normalisation, $x$, (Equation~\ref{eqn:Amati relation}) of 200 is the value used to make the light curves presented in the paper. It is set such that the IPC (pink point) sits on the EE Amati relation as determined by \citep{EE_amatirelation}. The normalisation of 30 was computed to test the spectral evolution and is explained below.

There has been limited investigation of where the EE should sit on the Amati relation when considered separately from the IPC. \cite{EE/IPCAmati} found the IPC to sit on the Amati relation for short GRBs whereas the EE sat on the long GRB relation for their sample. Both \cite{EE/IPCAmati} and \cite{EE_amatirelation} determined the IPC position to sit near the short Amati relation. \cite{Amati_060614} show that when considering the whole emission of EE burst 060614 it is consistent with the Amati relation for long GRBs. The whole emission is dominated by the longer EE tail. Therefore, this result suggests that the EE contribution is closer to the long GRBs rather than the shorts. \citet{Amati_060614} also consider the candidate long merger GRB 050724, for which they estimate the $E_p$ and $E_{\rm iso}$ for the soft component only. This is found to sit on the long Amati relation, consistent with the findings in \cite{EE/IPCAmati}. 

For a normalisation of 200, our EE point (blue) sits close to the IPC point (pink), consistent with the short GRB relation. However, with a lower normalisation of 30 our $E_p$ values are consistent with the long GRB population. This finding suggests a need for a time-varying $E_p$ -- $E_{\rm iso}$ normalisation in order to match observed trends, consistent with the possibility of time-varying physical conditions over the lifetime of the jet.

\section{Conclusions}
\label{sec:conclusions}

The aim of this work was to recreate the longer timescales of extended emission GRBs from a central engine that is active for only 2\,s (as expected for BNS mergers). We showed we can successfully lengthen the timescale of received emission with a shell model that incorporates high latitude emission and jet expansion. This consists of an initial, narrow jet of higher Lorentz factor ($\Gamma \sim$ 300) with a Gaussian lateral structure followed by a wider, tophat-like jet with lower Lorentz factor ($\Gamma \sim$ 9). Despite the lower Lorentz factors, we showed the jet remains optically thin for emission from enough lower Lorentz, wider shells to escape and produce the EE signal.

We compared our model light curve to the \emph{Swift} data of three representative EE GRBs (211211A, 211227A and 060614). Our canonical light curve can reproduce the observed features of EE bursts: an initial peak (IPC) followed by a broader and smoother EE plateau. We suggest that, within a select parameter range, the model is additionally able to reproduce standard short merger GRBs. 

We investigated the free parameters within the model, determining the most influential as the angular size of the jet core and the maximum Lorentz factor. We showed that the IPC is produced by a fast, narrow, Gaussian jet and the EE is the result of a slower and wider, tophat jet. 

We investigated the spectral evolution predicted by the model and compared to that measured in GRB 211211A. We found the evolution of the model spectrum over time to be slower than the observed rate, following a power-law evolution instead of the exponential decay measured in GRB 211211A \citet{Ben211211A}. However, the turnover time of the spectral evolution appeared to be consistent with the 211211A data. We suggest the slower spectral evolution may be an indication of other parameters changing inside the jet such as the particle number density or the magnetic field energy fraction. Future work will investigate the need for additional spectral evolution within the shells through the use of a time-varying normalisation when determining $E_p$ from the Amati relation. 

Our investigation shows that HLE and changing jet structure has the potential to resolve the tension between the expected accretion timescale and the observed emission duration. Our simplified model can successfully recreate the EE light curve phenomenon. Prospective upgrades to the treatment of spectral evolution and a connection to the physical processes within the jet hold the potential to deepen our understanding of the long merger process.


\section*{Acknowledgements}

IW is supported by the UKRI
Science and Technology Facilities Council (STFC).

BPG acknowledges support from STFC grant no.
ST/Y002253/1 and the Leverhulme Trust grant no. RPG-2024-117.

\section*{Data Availability}

The code used for the model is available upon reasonable request. The Swift data were downloaded from the UK Swift Science Data Centre (https://www.swift.ac.uk/).




\bibliographystyle{mnras}
\bibliography{example} 








\bsp	
\label{lastpage}
\end{document}